\documentclass[aps,prb,preprint,floatfix,groupedaddress,showpacs,longbibliography]{revtex4-1}

\usepackage{graphicx}
\usepackage{dcolumn}
\usepackage{amsmath,bm}
\usepackage{color}%
\usepackage{ulem}
\usepackage{multirow}
\usepackage{array} 
\usepackage{esvect}
\usepackage{xr}
\usepackage{braket}

\usepackage[utf8]{inputenc}
\usepackage{kotex}
\usepackage{kotex-logo}

\begin{document}


\title{Unveiling Giant Hidden Rashba Effects in Two-Dimensional Si$_2$Bi$_2$}

\author{Seungjun Lee}
\affiliation{Department of Physics and Research Institute for Basic Sciences, Kyung Hee University, Seoul, 02447, Korea}

\author{Young-Kyun Kwon}
\email[Corresponding author. E-mail: ]{ykkwon@khu.ac.kr}
\affiliation{Department of Physics and Research Institute for Basic Sciences, Kyung Hee University, Seoul, 02447, Korea}

\date{\today}
\begin{abstract}
Recently, it has been known that the hidden Rashba (R-2) effect in two-dimensional materials gives rise to a novel physical phenomenon called spin-layer locking (SLL).
However, not only has its underlying fundamental mechanism been unclear, but also there are only a few materials exhibiting weak SLL.
Here, through the first-principles density functional theory and model Hamiltonian calculation, we reveal that the R-2 SLL can be determined by the competition between the sublayer-sublayer interaction and the spin-orbit coupling (SOC), which is related to the Rashba strength. In addition, the orbital angular momentum distribution is another crucial point to realize the strong R-2 SLL.
We propose that a novel 2D material Si$_2$Bi$_2$ possesses an ideal condition for the strong R-2 SLL, whose Rashba strength is evaluated to be 2.16~eV{\AA}, which is the greatest value ever observed in 2D R-2 materials to the best of our knowledge.
Furthermore, we reveal that the interlayer interaction in a bilayer structure ensures R-2 states spatially farther apart, implying a potential application in spintronics.
\end{abstract}
%
\maketitle


\section{Introduction}
\label{Introduction}
The spin-orbit coupling (SOC) combined with an asymmetric crystal potential at surfaces or interfaces induces spin-polarized electronic states, called as Rashba (R-1) spin splitting.~\cite{{rashba1},{rashba2},{Manchon2015}}
The Rashba states exhibit unique band dispersion with spin-momentum locking, which can be described by
\begin{equation}
  H_\mathrm{R}=-\alpha_R\boldsymbol{\sigma}\times\mathbf{k}\cdot\hat{z}, \label{HR}
\end{equation}
where $\alpha_R$, $\boldsymbol{\sigma}$, and $\mathbf{k}$ represent a Rashba strength coefficient, a Pauli spin matrix vector, and a crystal momentum; and $\hat{z}$ indicates a direction of the local electric field induced by the asymmetric crystal potential. 
The unique physical properties of the Rashba states have been utilized to realize some crucial concepts in the spintronics,~\cite{{spin1},{spin2}} such as spin field transistor~\cite{{spintronics},{Koo1515}} and intrinsic spin Hall effects.~\cite{PhysRevLett.92.126603}

It has, however, been reported that the Rashba spin splitting is strongly affected by local orbital angular momentum (OAM) $\mathbf{L}$ in a system with a SOC,~\cite{{PhysRevLett.107.156803},{PhysRevB.85.195401},{PhysRevB.85.195402},{PhysRevB.88.205408}} which can be described by the orbital Rashba Hamiltonian
\begin{equation}
  H_\mathrm{L}=-\mathbf{p}\cdot\mathbf{E}=-\gamma\mathbf{k}\times\mathbf{E}\cdot\mathbf{L},
\label{HH}
\end{equation}
where $\mathbf{p}=\gamma\mathbf{L}\times\mathbf{k}$ is electric dipole moment produced by the asymmetric charge distribution,~\cite{PhysRevLett.107.156803} and $\gamma$ is a proportional coefficient. Since the Rashba effects from these two model Hamiltonians Eqs.~(\ref{HR}) and (\ref{HH}) may not be distinguished in band calculations, Eq.~(\ref{HR}) may be used to extract the Rashba strength $\alpha_R$ from the Rashba states.

Since the centrosymmetry guarantees the spin-degenerate electronic structures, only non-centrosymmetric system have been considered as candidates for the R-1 based spintronics applications.
Recently, however, new insights were suggested that local symmetry breaking may induce ``local Rashba'' (R-2) spin splitting even in centrosymmetric materials~\cite{{Zhang2014},{Riley2014}}.
In such systems, intriguingly, degenerate spin states protected by the centrosymmetry are spatially separated into each inversion partner, which can be experimentally detected by spin- and angle-resolved photoemission spectroscopy in both bulk and two-dimensional (2D) materials~\cite{{Riley2014},{MoS2},{Yao2017},{Santos-Cottin2016},{Wu2017}}.
Among materials exhibiting the R-2 effects, bulk systems are not suitable for utilizing the spatially-separated states because their localized spin states would be canceled out by their adjacent inversion partners.
In the van der Waals (vdW) 2D materials, on the other hand, opposite spins in the degenerate states can be split into the top and bottom layers (or atomic sub-layers).
Such a spatially-separated spin splitting is called spin-layer locking (SLL).~\cite{{Riley2014},{Yao2017},{MoS2}}

Even though a few experimental observations have shown clear evidences for the existence of the R-2 effects, the following important questions still remain unanswered. 1) Why do some R-2 materials exhibit parabolic band structure rather than the Rashba-like dispersion? 2) How can the R-2 SLL effect be distinguished from unavoidable substrate effects in the experiments?~\cite{PhysRevB.97.125434} 3) Why does the degree of spin segregation depend on the band index of an R-2 material?
In addition, some of R-2 materials display an energy gap between the upper and lower R-2 bands, which cannot be described by the conventional Rashba (R-1) model Hamiltonian given by Eq.~(\ref{HR}).
Therefore, a new model Hamiltonian is required to correctly describe the R-2 SLL.
Furthermore, to utilize the R-2 Rashba states in the spintronics applications, it is essentially demanded not only to understand the fundamental physics of the R-2 effects but also to search for 2D materials exhibiting strong R-2 SLL.

To answer and resolve these questions and issues, in this paper, we propose a novel vdW 2D material Si$_2$Bi$_2$ with strong R-2 SLL and explore the physical origins combining the first-principles density functional theory and a model Hamiltonian. We found that the strong SOC restricts wavefunction overlap between local inversion partner and enables the OAM to contribute to the band-selected Rashba effects, leading to the giant R-2 SLL. The Rashba strength of Si$_2$Bi$_2$ was calculated to be 2.16~eV{\AA}, which is the greatest value ever observed in 2D R-2 materials to the best of our knowledge.
In addition, we suggest that multilayer configuration may enhance the spatial segregation of spin splitting occurring only at the outermost surfaces, while diminishing almost completely at the inner ones due to the interlayer interactions. Such a stacking process eventually leads to the evolution from the R-2 to R-1 spin splitting.

\section{Results and discussion}
\label{Results}

\begin{figure}[t]
\includegraphics[width=1.0\columnwidth]{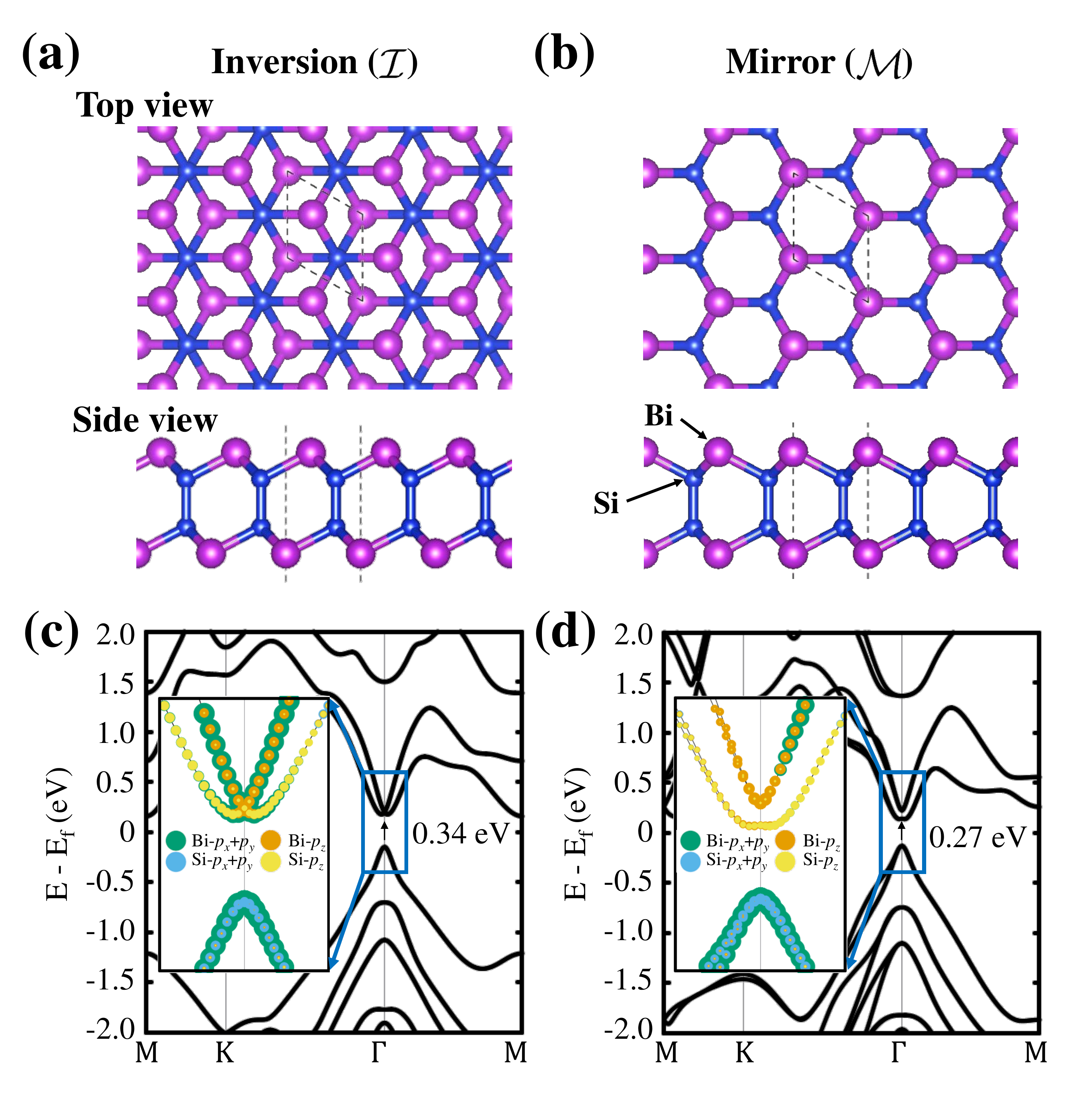}
\caption{Top and side views of the equilibrium crystal structures of Si$_2$Bi$_2$, where Si and Bi atoms are depicted by yellow and purple balls for two different phases with (a) inversion ($\mathcal{I}$) and (b) mirror ($\mathcal{M}$) symmetries, and (c, d) their respective electronic band structures, of which the orbital-resolved conduction and valence bands near the $\Gamma$ point are shown in insets, where each color is assigned to each projected orbital and the line thickness indicates the degree of the orbital contribution.
The calculated band gap values at the $\Gamma$ point are 0.27 and 0.34~eV, respectively.
\label{Structure}}
\end{figure}

Earlier studies reported that a group IV elements, X (X$=$C, Si, Ge, Sn, or Pb) can combine with a group V elements, Y (Y$=$N, P, As, Sb, or Bi) to form a stable layered compound X$_2$Y$_2$.~\cite{{45-prb},{45-nano},{Lee2019c}} Each layered compound can be classified into two groups by the crystal symmetries, one with the $\mathcal{I}$nversion symmetry ($p\bar{3}m1$) and the other the $\mathcal{M}$irror one ($p\bar{6}m2$), as shown in Figures~\ref{Structure} (a) and (b). It is noted that their space gruops of these two crystal phases of X$_2$Y$_2$ are essentially the same as those of the 1H and 1T phases of transition metal (M) dichalcogenides (A$_2$), MA$_2$, with the correspondences of X$_2$ and Y to M and A, respectively. Among various group IV-V compounds, we propose that Si$_2$Bi$_2$ becomes an exemplary material exhibiting strong R-2 type Rashba effects. It was found that both $\mathcal{I}$ and $\mathcal{M}$ phases of Si$_2$Bi$_2$ are stable with their energy difference of less than 3~meV/atom.~\cite{Lee2019c} 

Figures~\ref{Structure} (c) and (d) shows their corresponding electronic band structures calculated using the PBE XC functional, and their orbital-resolved ones in the respective insets, which will be discussed later.
Both $\mathcal{I}$- and $\mathcal{M}$-phases have a band gap of 0.34 and 0.27~eV at the $\Gamma$ point.~\footnote{it is noted that the $\mathcal{I}$-phase is an indirect band gap semiconductor with 0.31~eV since the conduction band at that M point is slightly lower that that at the $\Gamma$ point.}
The inversion symmetry in the $\mathcal{I}$ phase guarantees degenerate spin states, whereas the mirror symmetry in the $\mathcal{M}$ phase lifts the degeneracy.
Nevertheless, both phases revealed the Rashba-type splitting near their conduction band minima, but not near the valence band maxima in their band structures.
To verify that such a Rashba-type splitting is not a functional-dependent artifact, we also used the HSE06 XC functional, which yields Rashba-like band structures in both phases as shown in Fig.~S1 in the Supplementary Information (SI).
To get better understanding of these unusual Rashba effects, we first focus our discussion on the $\mathcal{I}$ phase with higher symmetric features and then on the $\mathcal{M}$ phase

\begin{figure}[t]
\includegraphics[width=0.75\columnwidth]{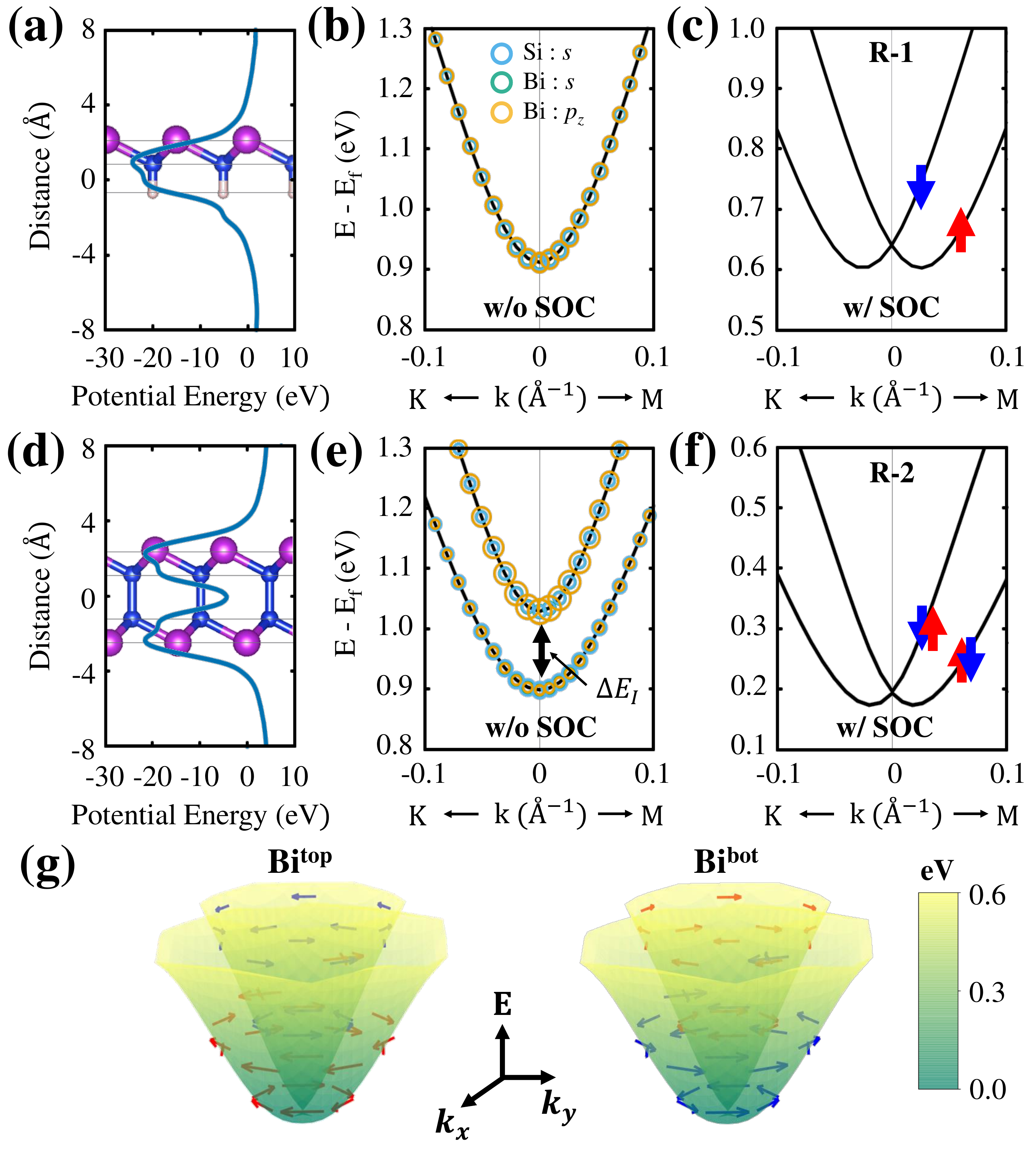}
\caption{ 
Plane-averaged electrostatic potentials and electronic structures near the conduction band minima of an artificially-constructed asymmetric 2D SiBi monolayer (a--c) and $\mathcal{I}$-Si$_2$Bi$_2$ (d--f). The respective structures, in which Si and Bi atoms are depicted by the purple and blue spheres, respectively, are overlaid in (a) and (d). The dangling bond of each Si atom in (a) is saturated with a hydrogen atom depicted with the white sphere.
The electronic band structures are shown without SOC in (b) and (e), and with SOC in (c) and (f). The latter two cases clearly show the Rashba band splittings, one with the R-1 and the other R-2. Local orbital contributions are displayed with skyblue, green, and yellow circles for the Si $s$, and the Bi $s$ and $p_z$ orbitals, respectively. The radius of circle indicates the degree of the orbital contribution. The $\Delta E_I$ in (e) indicates the splitting energy between CB1 and CB2 due to the interaction between the top and bottom SLs. The red and blue arrows denote spin states in (c) and (f).
(g) Spin textures spatially-resolved on the top (left) and bottom (right) Bi SLs or Bi$^\mathrm{top}$ and Bi$^\mathrm{bot}$ plotted on the Rashba bands shown in (f). Red (Blue) arrows represent the clockwise (counterclockwise) spin chirality and the colorbox indicates energy scale in eV relative to the bottom of the conduction band.
\label{Rashba}}
\end{figure}

To verify if the Bi-Si system indeed produces the Rashba splitting, as a first step, we constructed an artificial substrucure composed of Bi-Si monolayer with a broken inversion symmetry as shown in Fig.~\ref{Rashba} (a). The dangling bond of each Si atom in the 2D SiBi monolayer was saturated by a hydrogen atom. Due to the atomic arrangement of Bi-Si and the difference in their electron affinities, its strong asymmetric local potential shown in Fig.~\ref{Rashba} (a) induces an electric field perpendicular to its plane. 
We calculated its electronic band structure shown in Fig.~S4 in the SI. 
In the absence of SOC, as shown in Fig.~\ref{Rashba} (b), the lowest conduction band (CB1) exhibits spin-degenerate parabolic dispersion relation and is mainly formed by the Bi $p_z$ orbital.
With SOC turned-on, the $p_z$ orbital related to the in-plane OAM can give a significant contribution to the Rashba interaction due to the out-of-plane local electric field, as described in Eq.~(\ref{HH}).
Thus, the 2D SiBi monolayer exhibits a typical R-1 spin splitting as shown in the Fig.~\ref{Rashba} (c).

In the real $\mathcal{I}$-Si$_2$Bi$_2$ material, however, the underlying physics becomes much more complicated because SL-SL interaction also takes part in determining its electronic structure. Similar to Fig.~\ref{Rashba} (a), we also computed the local electrostatic potential of $\mathcal{I}$-Si$_2$Bi$_2$ displayed in Fig.~\ref{Rashba} (d). Due to the inversion symmetry, there is no net dipole moment. However, the potential profile clearly indicates that there are local dipole moments on the top and bottom SLs, which are oppositely oriented to each other. Were it not for both the SL-SL interaction and SOC, $\mathcal{I}$-Si$_2$Bi$_2$ should have had the four-fold degenerate conduction bands composed mainly of the $p_z$ orbitals with spin-up and -down of the top and bottom Bi SLs. Now only turning on SOC, the opposite local electric fields, each on each SL, should have induced the spatially-separated R-2 spin splitting, but the inversion symmetry still guarantees the spin-degeneracy, which are schematically summarized in Fig.~S5 in SI.

There is, however, an inevitable interaction between the top and bottom SLs, which lifts their four-fold degeneracy via the wavefunction overlap even without SOC, as shown in Fig.~\ref{Rashba} (e). The splitting energy $\Delta E_I$ due to such interaction was calculated to be 0.13~eV at the $\Gamma$ point. One could expect that SOC would split these bands further into two sets of Rashba bands, which is, however, contradictory to degeneracy guaranteed by the inversion symmetry.
It was instead surprisingly found that turning on SOC converted two separated doubly-degenerate parabolic bands (Fig.~\ref{Rashba} (e)) to almost perfect and doubly-degenerate Rashba bands, as shown in Fig.~\ref{Rashba} (f), essentially the same as those expected in the case without the SL-SL interaction as described above and in 
Fig.~S5 in SI. To confirm that such bands are the hidden 
Rashba bands, we explored the doubly-degenerate Rashba bands (Fig.~\ref{Rashba} (f)) by computing the spatially-resolved spin textures~\footnote{\label{note2}For a given $(n,\mathbf{k})$, the spin polarization is the expectation value of the spin operator, that is, $\boldsymbol{\mu}_{n,\mathbf{k}}\propto\braket{\psi_{n,\mathbf{k}}|\boldsymbol{\sigma}|\psi_{n,\mathbf{k}}}$, where $\boldsymbol{\sigma}$ is the Pauli spin matrix vector. Its projection on each atom $\alpha$, $\boldsymbol{\mu}_{n,\mathbf{k}}^\alpha$, was obtained by expanding $\boldsymbol{\mu}_{n,\mathbf{k}}$ in terms of the spherical harmonics $\ket{Y_{lm}^\alpha}$ with the orbital angular momentum $(l,m)$ of the $\alpha$ atom, or \unexpanded{$\boldsymbol{\mu}_{n,\mathbf{k}}^\alpha\propto\sum_{l,m}\braket{\psi_{n,\mathbf{k}}|\boldsymbol{\sigma}|Y_{lm}^\alpha}\braket{Y_{lm}^\alpha|\psi_{n,\mathbf{k}}}$}.} on the top (Bi$^\mathrm{top}$) and bottom (Bi$^\mathrm{bot}$) Bi atom SLs shown in Fig.~\ref{Rashba} (g). The spin map on each layer exhibits the opposite spin chiralities on the inner and outer Rashba bands, as usually observed in noncentrosymmetric systems. Even more intriguingly, these spin chiralities are spatially coupled to the layers. The spins of Bi$^\mathrm{top}$ on the inner band rotate in one way (e.g., counterclockwise, blue arrows in the left image), whereas those of Bi$^\mathrm{bot}$ do in the other way (clockwise, red arrows on the right image). On the outer band, their corresponding spins rotate the other way around. (red arrows on the left and blue arrows on the right images) This observation clearly reveals the strong R-2 SLL phenomenon. Furthermore, it is worthwhile to note the relation between the local symmetry breaking and the orbital polarization described in Eq.~(\ref{HH}). We investigated the OAM distributions~\cite{CHPark} of both the 2D SiBi monolayer and $\mathcal{I}$-Si$_2$Bi$_2$ as described in Sect.~\ref{Computational}, and found that the $\mathcal{I}$-Si$_2$Bi$_2$ exhibits the ``hidden'' orbital polarization as well as the R-2 SLL induced by the local symmetry breaking. For detailed description, see Note~S1 with Figs.~S2 and S3 in SI.

\begin{figure*}[t]
\includegraphics[width=1.0\textwidth]{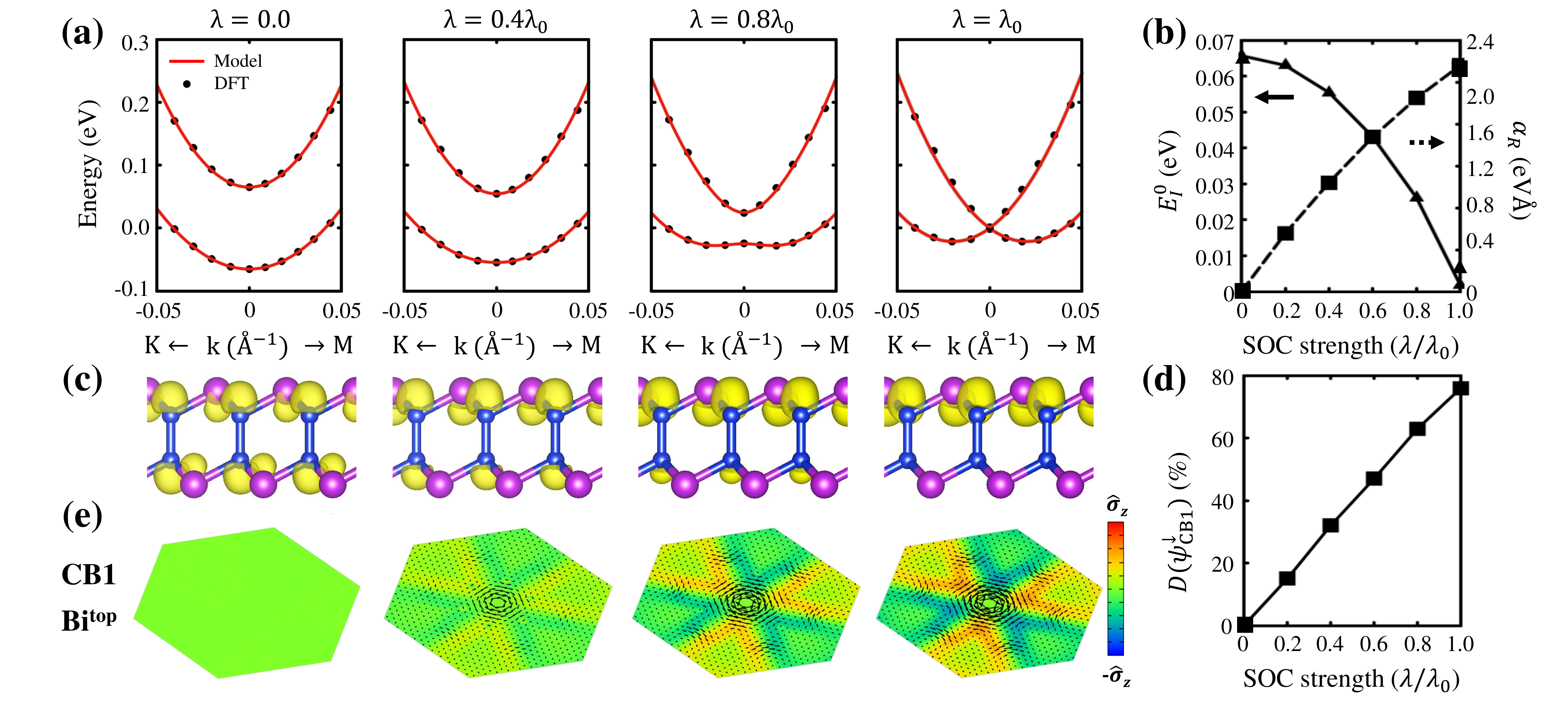}
\caption{Evolution of the electronic structure with the SOC strength $\lambda/\lambda_0\in(0,1)$, where $\lambda_0$ is the real SOC strength: (a) two lowest conduction bands near the $\Gamma$ point.
The black dots show the energy eigenvalues calculated by DFT, which were fitted by the model defined in the Eq.~(\ref{eigen}), plotted with red solid lines. 
(b) $\lambda$-dependence of the interaction energy ($E_I^0$) and the Rashba strength ($\alpha_R$) fitted in (a).
(c) $|\psi^\downarrow_\mathrm{CB1}|^2$, spin-resolved wavefunction squared, calucated at $\mathbf{k_{\Gamma-M}}=(0.015,0) (2\pi/a)$ near the $\Gamma$ point. 
(d) Degree of wavefunction segregation D($\psi$) defined in Eq.~(\ref{DWS}) evaluated from (c) as a function of $\lambda$.
(e) Spatially-resolved spin maps on the top Bi atom layer plotted in the first Brillouin zone for the lowest conduction band. The size of black arrows and different colors indicate the in-plane and out-of-plane spin components. 
\label{soc_variation}}
\end{figure*}

In view of previous results observed in other 2D R-2 materials, such as PtSe$_2$~\cite{Yao2017} or bilayer WSe$_2$~\cite{Riley2014}, which have revealed the SLL phenomena, but still with parabolic bands similar to those shown in Fig.~\ref{Rashba} (e), our hidden Rashba bands shown in Fig.~\ref{Rashba} (f) are exceptionally unusual since they look like the Rashba-like bands shown in Fig.~\ref{Rashba} (c).
To answer what causes such distinction, we examined the pathway from the parabolic bands (Fig.~\ref{Rashba} (e)) to the Rashba-like ones (Fig.~\ref{Rashba} (f)) while manipulating the SOC strength $\lambda/\lambda_0\in(0,1)$, where $\lambda_0$ is the real SOC strength of our $\mathcal{I}$-Bi$_2$Si$_2$ system.  
As $\lambda$ increases, two split bands tend to form a Rashba-like bands through continuous change as shown in Fig.~\ref{soc_variation} (a).
This result indicates that the competition between the SL-SL interaction ($\Delta E_I$) and the Rashba strength ($\alpha_R$) determines the shape of electronic bands.

To understand such competition quantitatively, we devised a simplest model Hamiltonian represented by four minimal basis vectors that describes the SL-SL interaction and Rashba splitting. For more detailed 
description of our model Hamiltonian, see Note~S2. in SI.
From the model Hamiltonian, we obtained two doubly-degenerate bands 
\begin{equation}
\label{eigen}
E_{\pm}=\frac{\hbar^2k^2}{2m^*}\pm\sqrt{\left(\alpha_R k\right)^2+\left(E_I^0+E_I^1k^2\right)^2},
\end{equation}
%
%
guaranteed by inversion symmetry. Here $m^*$ is the effective mass, $E_I^0$ and $E_I^1$ the SL-SL interaction coefficients. To discover how to compete the SL-SL interaction with Rashba spin splitting, the DFT bands (black dots) were fitted to the model bands (red lines) given in Eq.~(\ref{eigen}) resulting in almost perfect agreement as shown Fig.~\ref{soc_variation} (a). 
The fitted parameters $E_I^0$ and $\alpha_R$ as a function of $\lambda$ are shown in Fig.~\ref{soc_variation} (b). As expected, SOC weakens the SL-SL interaction, but strengthens $\alpha_R$, which was calculated to be 2.16~eV{\AA} at $\lambda=\lambda_0$.
This value is much larger than those observed in metal surfaces, for example Au(111) (0.33~eV\AA),~\cite{PhysRevLett.77.3419} Bi(111) (0.55~eV\AA),~\cite{PhysRevLett.93.046403} as well as other materials exhibiting the R-2 SOC such as BaNiS$_2$ (0.24~eV\AA),~\cite{Yuan2019} 
and is also comparable with conventional giant Rashba system, such as hybrid perovskites (1.6~eV\AA),~\cite{perov} BiSb monolayer (2.3~eV\AA),~\cite{BiSb} or BiTeI (3.8~eV\AA),~\cite{Ishizaka2011}.
Therefore, we may classify our system into the first ``giant hidden'' Rashba material.

This result was further confirmed by $|\psi^\downarrow_\mathrm{CB1}(\mathrm{r})|^2$, obtained from the spin-resolved wavefunction yielded near the $\Gamma$ point. As shown in Fig.~\ref{soc_variation} (c), it evolves from an even distribution on both SLs at $\lambda=0$ toward a complete spatial segregation at $\lambda=\lambda_0$, which is quantified by $D(\psi)$ defined by Eq.~(\ref{DWS}), shown in Fig.~\ref{soc_variation} (d).
For every $\lambda$ value, we also reckoned the spatially-resolved spin map on the Bi$^\mathrm{top}$ SL to verify the degree of the SLL, which shown in Fig.~\ref{soc_variation} (e). 
We emphasize that no sharp phase transition was observed and thus even for $\lambda<\lambda_0$, the system exhibits the SLL while maintaining two parabolic bands due to appreciable SL-SL interaction. 
When $\lambda$ becomes $\lambda_0$ eventually, all three features clearly reveal complete Rashba-like bands, wavefunction segregation and SLL implying that our system, 2D $\mathcal{I}$-Si$_2$Bi$_2$ possesses vastly strong SOC minimizing the SL-SL interaction.
We further notice that when $\lambda$ becomes larger than $\lambda_0$, energy gap between CB1 and CB2 reopens, while wavefunctions are still strongly segregated and R-2 SLL is also vivid, as shown in Fig.~S6 in SI.

On the other hand, we noticed that there is no Rashba spin splitting at the highest valence band (VB1) unlike at CB1, perusing the band structures shown in Fig.~\ref{Structure} (c). As shown in the inset of Fig.~\ref{Structure} (c) and Fig.~S7 (a), there is nearly no $p_z$ orbital contribution at VB1, resulting in no OAM distribution to produce Rashba spin splitting even with strong SOC, which is clear from Eq.~(\ref{HH}). This non Rashba feature observed in the VB1 was further confirmed by the spin texture and the wavefunction segregation computed 
on the VB1 shown in Fig.~S7 (b) and (c) in SI. 
Therefore, to utilize an R-2 material in the spintronics application, its hidden Rashba SLL should be induced by the bands near the Fermi level, which possess the OAM perpendicular to the local electric field.

At this time, it is worth mentioning that the spin splitting was also observed in $\mathcal{M}$-Si$_2$Bi$_2$ with the broken inversion symmetry, which additionally lifts the degeneracy protected in the $\mathcal{I}$-counterpart and guarantees the Dresselhaus spin splitting.~\cite{Dresselhaus}
Intriguingly, we also observed a strong SLL in a few lowest conduction
bands near the $\Gamma$ point, as shown in Fig.~S8 in SI. 
Such strong SLL is also attributed to the OAM similar in its inversion counterpart. Here, we again emphasize that the OAM is an important factor to determine the SLL.

\begin{figure}[t]
\includegraphics[width=1.0\columnwidth]{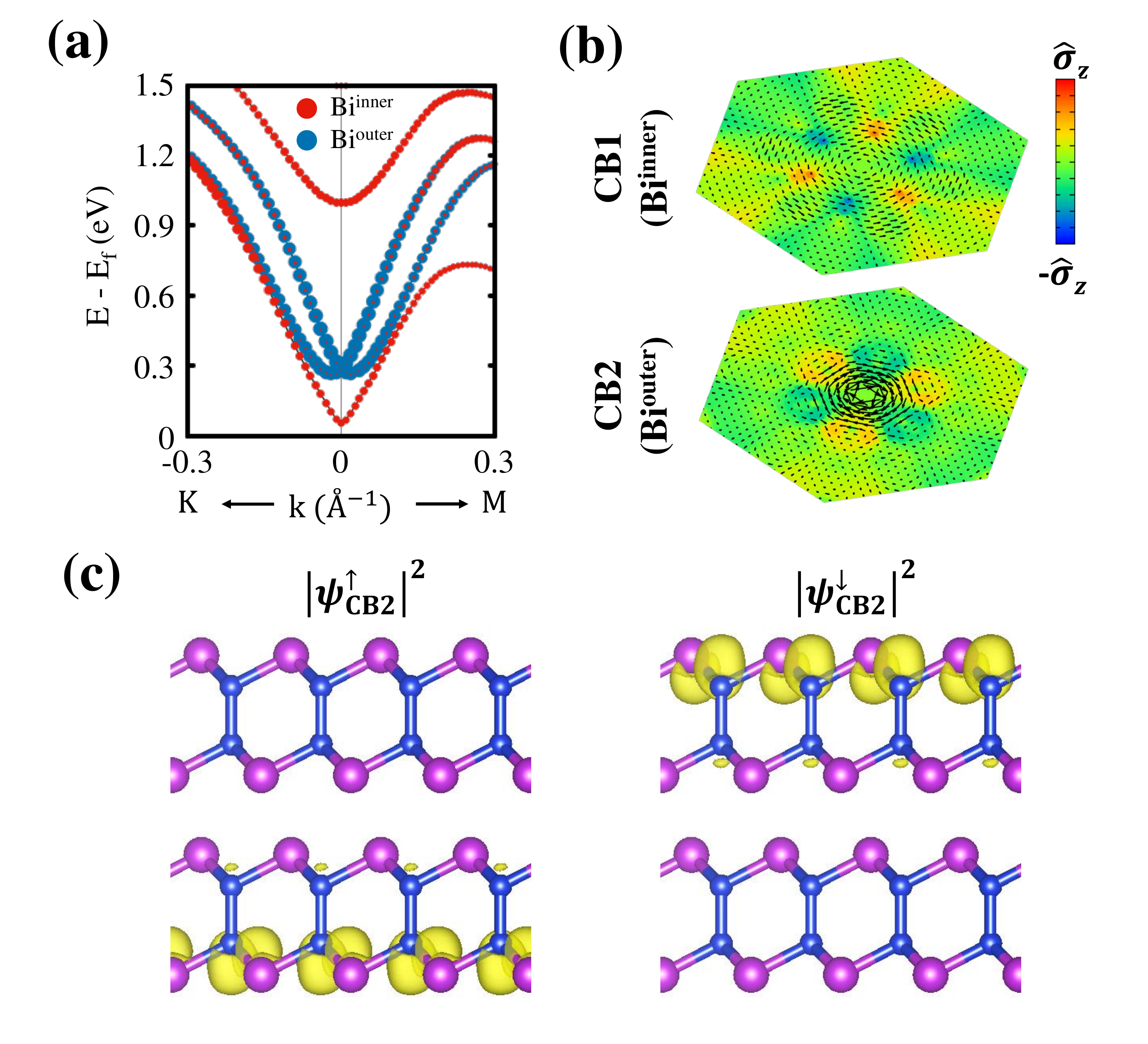}
\caption{
(a) Four lowest conduction bands (CB1 through CB4 in ascending order in energy) of a $\mathcal{I}$-Si$_2$Bi$_2$ bilayer spatially-resolved on the inner Bi SLs, Bi$^\mathrm{inner}$ (red circles) and on the outer ones, Bi$^\mathrm{outer}$ (blue circles). The size of circles indicates the amount of their contributions, which are well separated especially near the $\Gamma$ point. 
(b) Spin textures computed from the CB1 and CB2 localized on Bi$^\mathrm{inner}$ and Bi$^\mathrm{outer}$, respectively. 
The size of black arrows and different colors indicate respectively the in-plane and out-of-plane spin components.
(c) Spatially-projected spinor wavefunction squared, $|\psi^\sigma_\mathrm{CB2}|^2$, with $\sigma={\uparrow}$ and ${\downarrow}$ for the CB2 band evaluated at $k=0.03\pi/a$ along the $\Gamma-$M line close to the $\Gamma$ point.
\label{bilayer}}
\end{figure}

Since 2D materials usually form multilayers rather than monolayers, it is also of importance to understand the effect of the interlayer- or vdW interaction on the SLL phenomenon. To do this, we constructed a bilayer of $\mathcal{I}$-Si$_2$Bi$_2$ with ``AA'' stacking
which is still maintains the inversion symmetry, and investigated its  electronic properties. Figure~\ref{bilayer} (a) shows its four lowest conduction bands (CB1 through CB4 in ascending order in energy) represented by the projections on the inner Bi (Bi$^\mathrm{inner}$) SLs in red circles and on their outer (Bi$^\mathrm{outer}$) counterparts in blue circles. The widely-split two red bands (CB1 and CB4) are contributed mostly by the Bi$^\mathrm{inner}$ SLs, whereas the Rashba-like blue bands (CB2 and CB3) mostly by the Bi$^\mathrm{outer}$ ones. Were it not for the vdW interaction, two identical monolayers should have shown exactly the same band structures as shown in Fig.~\ref{Rashba} (f) with every band four-fold degenerated. One can, however, easily expect that the Bi$^\mathrm{inner}$ SLs are directly affected by the vdW interaction, whereas the Bi$^\mathrm{outer}$ SLs are intact. Thus it can be explained that the vdW interaction caused the band repulsion between the CB1 and CB4, while the CB2 and CB3 localized at two opposite Bi$^\mathrm{outer}$ SLs preserve the SLL induced in each isolated monolayer. 

Our explanation becomes even clearer from the spatially-resolved spin textures and spinor wavefunctions. Spin texture of the Bi$^\mathrm{outer}$ SLs evaluated on the CB2 does show much stronger spin chirality implying stronger SLL than that of the Bi$^\mathrm{inner}$ SLs on the CB1, where R-2 Rashba effects are suppressed, as shown in Fig.~\ref{bilayer} (b). 
To verify if the SLL in the bilayer is indeed from the CB2 and CB3, we represented the spinor wavefunctions squared, $\left|\psi_n^\sigma\right|^2$ in real space. In Fig.~\ref{bilayer} (c), those for $\sigma={\uparrow}$ and $\downarrow$ calculated on the $n=$~CB2 clearly display spatially-segregated spin states demonstrating the strong SLL. It is worthy of mentioning that each band is still doubly degenerate since the bilayer configuration also possesses the inversion symmetry. In other word, it is the interlayer interaction that removes the R-2 effect at the inner surfaces, but the Rashba states survive only at the outer surfaces of the bilayer. This spatial segregation would be utilized in some spintronics applications since one could control the spin behaviors only on the top surface without being influenced by those on the bottom one. 
We also noticed that when the layered R-2 materials become a bulk structure, Rashba spin splittings inside bulk region may be removed as seen in the bilayer, and only the surface Rashba states survive, implying that the R-2 SLL automatically changes to the R-1 spin splitting.
It is, therefore, the local symmetry breaking that is the physical origin causing not only the R-2 SLL, but also R-1 effects.

\section{Conclusion}
\label{Conclusion}
Using first-principles density functional theory, we predicted a new 2D material, which is layered Si$_2$Bi$_2$ exhibiting the giant R-2 SLL.
To understand an underlying physical origin of R-2 SLL, we performed first-principles calculations as well as solved a devised model Hamiltonian to describe the R-2 SLL.
Through this model calculation, we found that there is a competition between the SOC and SL-SL interaction to reveal the R-2 SLL. As the former, as it increases, weakens the latter and strengthens $\alpha_R$ leading to the giant hidden Rashba spin splitting. Furthermore we found that the R-2 SLL is also closely related to the OAM distribution. 
The Rashba strength in Si$_2$Bi$_2$ was calculated to be 2.16~eV{\AA}, which is the greatest value ever observed in R-2 materials to the best of our knowledge. We also revealed from a bilayer case that the R-2 SLL can be removed at the inner surfaces due to the interlayer interaction, but remained spatially farther apart at the outer surfaces. This eventually leads to a conclusion that the R-1 effect is also originated from the same local symmetry breaking causing the R-2 SLL.
Our findings may not only uncover the fundamental physics of R-2 SLL, but also provide a guidance for searching novel R-2 materials.

\section{Methods}
\label{Computational}
To understand the underlying physics of the R-2 SLL in 2D Si$_2$Bi$_2$,
we performed first-principles calculations based on density functional
theory~\cite{Kohn1965} as implemented in Vienna \textit{ab initio}
simulation package (VASP)~\cite{Kresse1996}. The electronic wavefunctions were expanded by plane wave basis with kinetic energy cutoff of 500~eV. 
We employed the projector-augmented wave pseudopotentials
~\cite{{Blochl1994},{Kresse1999}} 
to describe the valence electrons, and treated exchange-correlation (XC) functional within the generalized gradient approximation of 
Perdew-Burke-Ernzerhof (PBE)~\cite{Perdew1996} with noncollinear spin polarization.~\cite{PhysRevB.93.224425} To rule out any functional-related artifacts, we verified our PBE-based results using hybrid functional (HSE06).~\cite{HSE06}
For bilayer calculations, in which interlayer interaction cannot be neglected, Grimme-D2 Van der Waals correction~\cite{grimme-d2} was added.
To mimic 2D layered structure in periodic cells, we included a sufficiently large vacuum region in-between neighboring cells along the out-of-plane direction.
The Brillouin zone (BZ) of each structure was sampled using a $30{\times}30{\times}1$ $k$-point mesh according to the Monkhost-Pack scheme.~\cite{MP}

To describe and visualize the R-2 SLL, we included spin-orbit interaction in the all calculation.
The SOC is described by an additional Hamiltonian 
\begin{equation}
H_\mathrm{SOC}^{\alpha\beta}=\frac{\hbar^2}{(2m_ec)^2}\frac{K(r)}{r}\frac{dV(r)}{dr}\boldsymbol{\sigma}^{\alpha\beta}\cdot\mathbf{L},
\label{SOC}
\end{equation}
where $\alpha$ and $\beta$ indicate spin-up and down components of the spinor wave function; $\boldsymbol{\sigma}$ and $\mathbf{L}$ are Pauli spin matrices and angular momentum operator; $V(r)$ is the spherical part of the effective all electron potential
within the PAW sphere; and 
\[
K(r)=\left(1-\frac{V(r)}{(2m_ec)^2}\right)^{-2},
\]
as explained in Refs. [\onlinecite{PhysRevB.93.224425}] and [\onlinecite{socH}].
To explore the effect of the SOC strength, we introduced an artificial parameter $\lambda$ which scales Eq.~(\ref{SOC}) as 
\[H_\mathrm{SOC}^{\alpha\beta}(\lambda)=\frac{\lambda}{\lambda_0}H_\mathrm{SOC}^{\alpha\beta}. \]
When $\lambda=\lambda_0$, it becomes the full SOC Hamiltonian given in Eq.~(\ref{SOC}).

The momentum-resolved spinor wavefunctions were evaluated by projecting the two-component spinor
%
\[
  \ket{\psi_{n\mathbf{k}}} = \left( 
  \begin{array}{c}
    \psi^{\uparrow}_{n\mathbf{k}} \\
    \psi^{\downarrow}_{n\mathbf{k}}
  \end{array}
  \right)
\]
into spherical harmonics $Y^{\alpha}_{lm}$ centered at ion index $\alpha$ with angular momentum quantum numbers $(l,m)$. Here $n$ and $\mathbf{k}$ are band index and crystal momentum, and the arrows $\uparrow$ and $\downarrow$ represent spin up and down. Such projected components were further manipulated to understand the contribution of each orbital angular momentum to the band structures and to generate the atom-resolved spin texture map.

To verify the SLL in our system, we quantify the spatial spin separation by introducing the degree of wavefucntion segregation (DWS) $D(\psi^\sigma_\mathbf{k})$ defined as~\cite{Yuan2019},
\begin{equation}
D(\psi^\sigma_{n\mathbf{k}})=\left\lvert\frac
{P_{\psi^\sigma_{n\mathbf{k}}}(S_\alpha) - P_{\psi^\sigma_{n\mathbf{k}}}(S_\beta)}
{P_{\psi^\sigma_{n\mathbf{k}}}(S_\alpha) + P_{\psi^\sigma_{n\mathbf{k}}}(S_\beta)}
\right\rvert,
\label{DWS}
\end{equation}
with
\begin{equation}
P_{\psi^\sigma_{n\mathbf{k}}}(S_{i})=\int_{\Omega\in S_{i}}
\left\lvert\psi^\sigma_{n\mathbf{k}}(\mathbf{r})\right\rvert^2 d^3 \mathbf{r},
\end{equation}
where $\sigma=\uparrow$ and $\downarrow$, $n$ is band index, and $S_i$ indicates the real space sector for the upper Bi-Si ($i=\alpha$) or lower Si-Bi ($i=\beta$) SL. $P_{\psi^\uparrow_\mathbf{k}}(S_{\alpha})$, for example, represents the wavefunction $\psi^\uparrow_{n,\mathbf{k}}$ localized on the upper SL sector $S_\alpha$.

\acknowledgments
We gratefully acknowledge financial support from the Korean government through the National Research Foundation (NRF) of Korea
(No. 2019R1A2C1005417). Some portion of our computational work was
done using the resources of the KISTI Supercomputing Center
(KSC-2018-CHA-0052 and KSC-2020-CRE-0011).

\bibliographystyle{apsrev}
\bibliography{my} 

\end{document}



Supplementary Information \\

\title{Unveiling Giant Hidden Rashba Effects in Two-Dimensional Si$_2$Bi$_2$}

\author{Seungjun Lee}
\affiliation{Department of Physics and Research Institute for Basic Sciences, Kyung Hee University, Seoul, 02447, Korea}

\author{Young-Kyun Kwon}
\email[Corresponding author. E-mail: ]{ykkwon@khu.ac.kr}
\affiliation{Department of Physics and Research Institute for Basic Sciences, Kyung Hee University, Seoul, 02447, Korea}

\date{\today}
\maketitle

\begin{figure}[p]
\includegraphics[width=1.0\columnwidth]{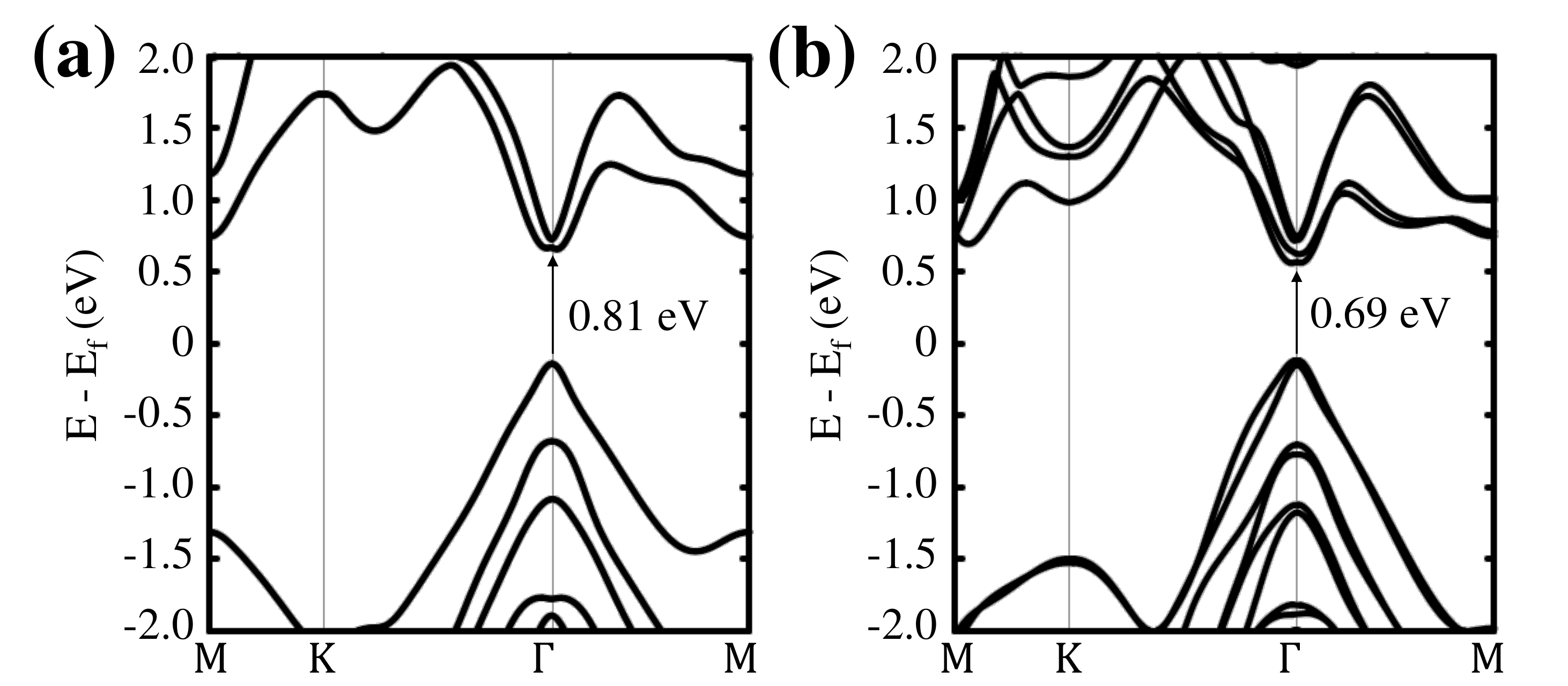}
\caption{Electronic band structures of (a) $\mathcal{I}$- and (b) $\mathcal{M}$-Si$_2$Bi$_2$ calculated by HSE06 functional.
The calculated band gap values at the $\Gamma$ point are 0.81 and 0.69 eV, respectively.}
\label{HSE_band}
\end{figure}

%
\section{Hidden orbital polarization}
\label{OAM}
\begin{figure}
\includegraphics[width=1.0\columnwidth]{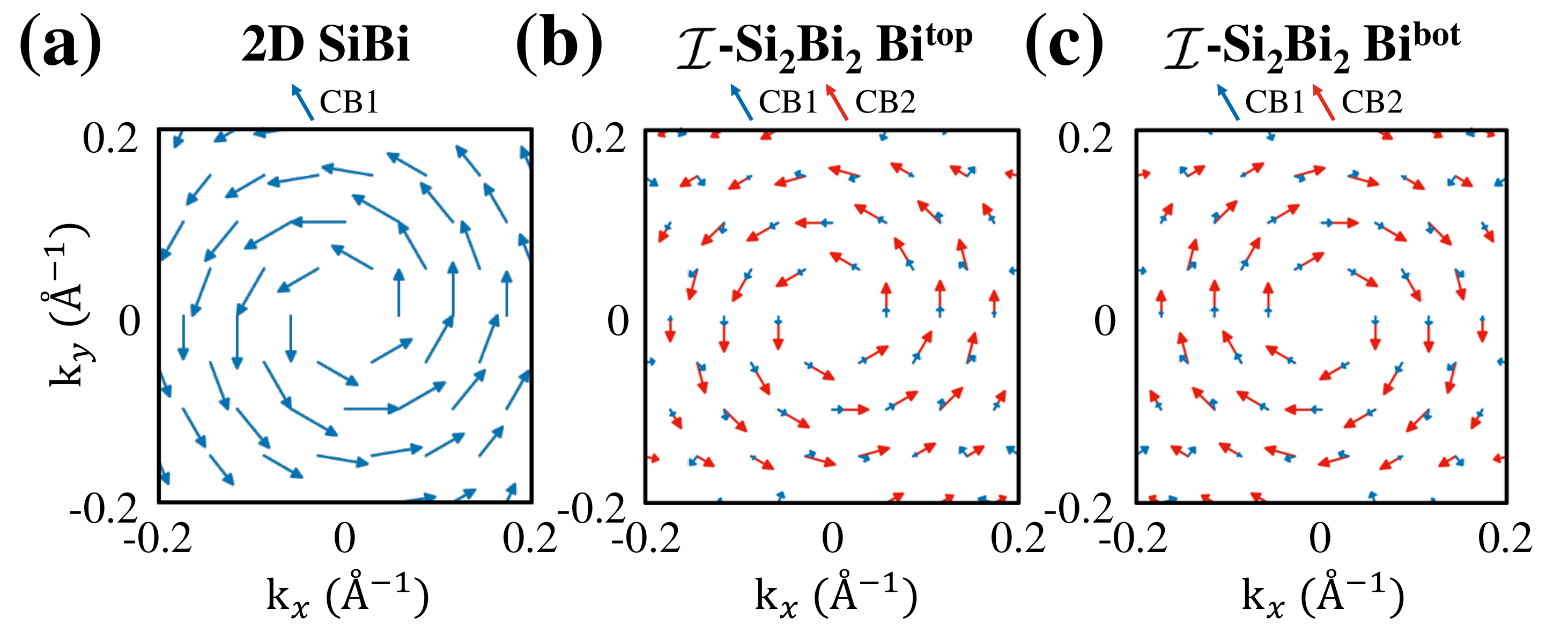}
\caption{The orbital polarization maps of (a) the 2D SiBi monolayer, and from (b) the top and (c) bottom Bi atoms of $\mathcal{I}$-Si$_2$Bi$_2$, near the $\Gamma$ point in the absence of SOC. The blue and red arrows in each map indicate the orbital polarization evaluated on the lowest conduction band (CB1) and second lowest conduction band (CB2), respectively.
\label{OAM1}}
\end{figure}

To check what the orbital polarization is for each band state, we visualized the orbital polarization by evaluating $\braket{\psi_{n,\mathbf{k}}|\mathbf{L}^\alpha}{\psi_{n,\mathbf{k}}}$,
the expectation value of the orbital angular momentum (OAM) operator defined by
\[
  L_i^\alpha=-i\hbar \sum_{j,k}\epsilon_{ijk}\ket{p^{\alpha}_{j}}\bra{p^{\alpha}_{k}},
\]
where $i,j,k=x,y,z$; $\epsilon_{ijk}$ is the Levi-Civita symbol; and $\ket{p^{\alpha}_{j}}$ is the $p_{j}$ orbital state at atom site $\alpha$.~\cite{CHPark}
This is essentially the same procedure as done for the spin polarization, which we described in the note [24] in the Reference Section in the main text.

Figure~\ref{OAM1} shows the orbital polarization maps in the absence of SOC for the two structures shown Fig.~2 in the main text. The lowest conduction band (CB1) of 2D SiBi monolayer, which is doubly-degenerate without SOC, is contributed mostly by the Bi $p_z$ orbitals, as shown in Fig.~2 (b) in the main text, resulting in significant OAM contribution. Thus the map of its OAM distribution (Fig.~\ref{OAM1} (a)) clearly exhibits the momentum-locked orbital polarization, which is exactly what we expected from the orbital Rashba Hamiltonian given in Eq.~(2) in the main text. For $\mathcal{I}$-Si$_2$Bi$_2$, on the other hand, the sublayer-sublayer interaction splits four-fold degenerate conduction bands into two doubly-degenerate bands, CB1 and the second lowest conduction band (CB2), in which the Bi $p_z$ orbital contribution is small and large, respectively, as shown in Fig.~2 (e) in the main text. Therefore, the orbital polarization is more vivid in CB2 (red arrows) than in CB1 (blue arrows), regardless of either the top or bottom Bi sublayer, but two sublayers exhibit their OAM distributions with the opposite chirality, as shown in Fig.~\ref{OAM1} (b) and (c), indicating that the total OAM of the whole system on each band will disappear. Therefore, $\mathcal{I}$-Si$_2$Bi$_2$ exhibits ``hidden'' orbital polarization or orbital-layer locking even without SOC.

\begin{figure}
\includegraphics[width=1.0\columnwidth]{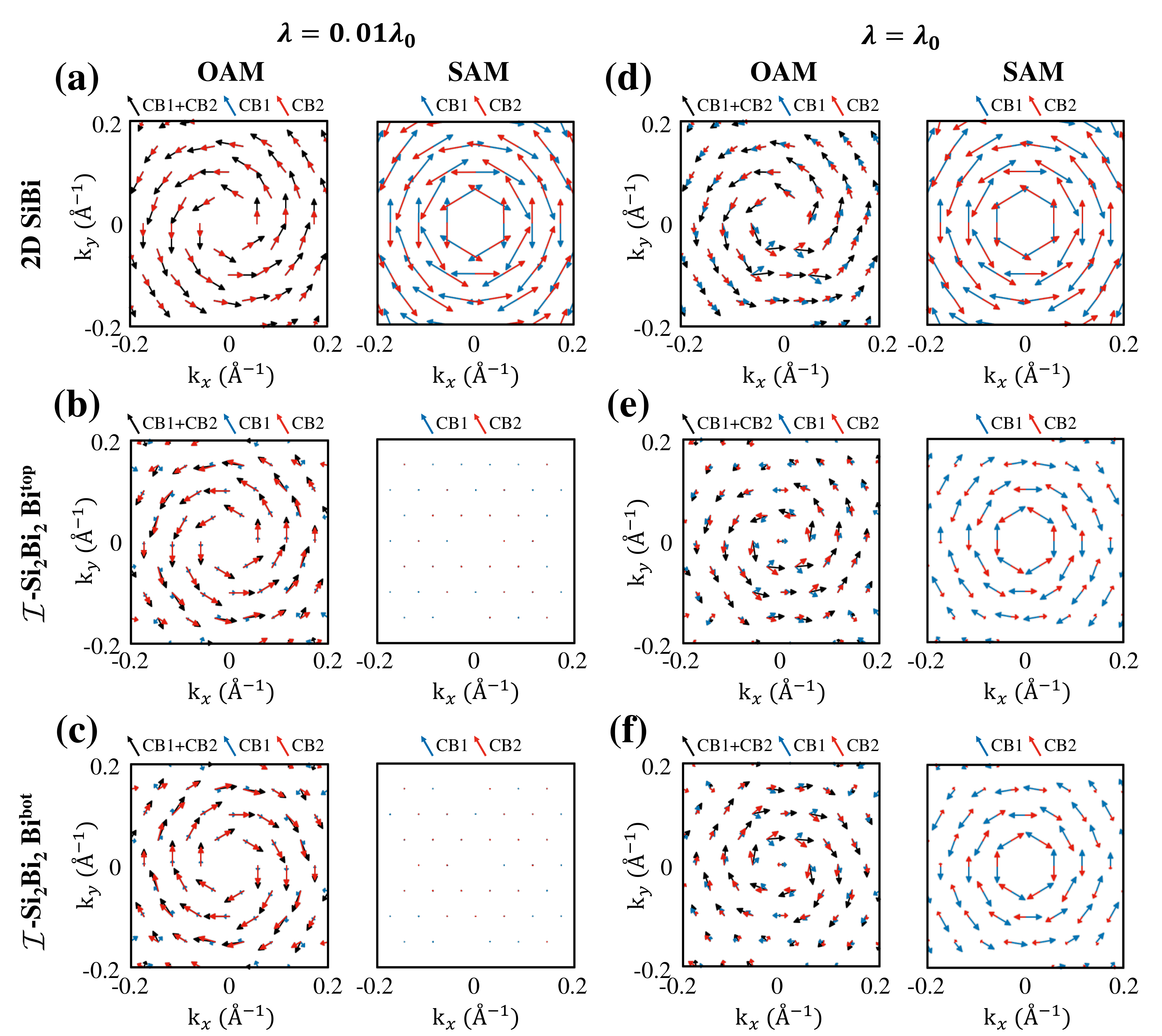}
\caption{The orbital angular momentum (OAM) and spin angular momentum (SAM) distributions when SOC is weak ($\lambda=0.01\lambda_0$) (a--c)  and strong ($\lambda=\lambda_0$) (d--f). The maps in the first low (a, d) display the orbital and spin polarizations of the 2D SiBi, and those in the second (b, e) and third lows (c, f) represent the polarizations projected on the top and bottom Bi atoms of $\mathcal{I}$-Si$_2$Bi$_2$, respectively. The maps in (a--c) are essentially the same as those in Fig.~\ref{OAM1} due to weak SOC, except for small band splitting. The blue and red arrows in each map indicate the polarization evaluated in CB1 and CB2, respectively, and the black arrows represent the sum of orbital polarizations in CB1 and CB2.}
\label{OAM2}
\end{figure}

With SOC on, we also computed the orbital polarizations of not only 2D SiBi monolayer, but also $\mathcal{I}$-Si$_2$Bi$_2$. Figure below (Fig.~\ref{OAM2}) shows their calculated OAM distributions as well as their spin angular momentum (SAM) distributions for comparison, when SOC is weak (a--c) and strong (d--f). The OAM maps for weak SOC are essentially the same as those in Fig.~\ref{OAM1} without SOC, except that the monolayer case revealed the band-split polarizations due to weak SOC. Since the 2D SiBi does not have inversion symmetry, even weak SOC ($\lambda=0.01\lambda_0$) split its spin-degenerate conduction band (CB1) into two spin polarized states, CB1 and CB2, even though their energy difference is negligibly small. Its SAM map displays that the spin polarizations on CB1 and CB2 are of the same magnitude with the opposite chirality, whereas the orbital polarizations on CB1 and CB2 are not only of the same magnitude, but also of the same chirality. Both maps clearly display the spin- and orbital-momentum locking on respective bands. For $\mathcal{I}$-Si$_2$Bi$_2$, the SAM maps projected on the top and bottom Bi atoms do not show any noticeable spin polarization as shown in Fig.~\ref{OAM2} (b) and (c), since the weak SOC cannot lift any one of the spin-degenerate CB1 and CB2, which are significantly separated by the sublayer-sublayer interaction and guaranteed by inversion symmetry. However, the spatially-projected OAM mpas clearly show hidden orbtial polarizations or orbtial-layer locking as mentioned above for the case without SOC.

Figure~\ref{OAM2} (d--f) shows the OAM and SAM maps with strong SOC. For the monolayer structure, its spin polarizations on CB1 and CB2 remain strong with a smalll difference in magnitude at each $k$-point, due to the Rashba splitting, and its OAM map still shows the momentum-locked polarization, which is not perfectly chiral though. For $\mathcal{I}$-Si$_2$Bi$_2$, its spatially-projected SAM maps clearly displays the hidden spin polarizations or spin-layer locking on both CB1 and CB2 induced by local broken inversion symmetry together with strong SOC. In its spatially-projected OAM maps, it was shown that the chirality of the orbital polarization (CB1+CB2) on one Bi atom sublayer is opposite to that on the other sublayer similar to those with weak SOC, however the chirality of the polarization on one sublayer is not completely opposite to that on the other on both CB1 and CB2 resulting in small, but non-zero polarization on each $k$-point.

\section{Model Hamiltonian}
\label{model}
To better describe the hidden Rashba (R-2) effect, we devised a simple model Hamiltonian, which explicitly takes into account not only Rashba splitting, but also the sublayer-sublayer (SL-SL) interaction. Our model Hamiltonian $\mathcal{H}$ can be decomposed into
\[
  \mathcal{H}=\mathcal{H}_0+\mathcal{H}_R+\mathcal{H}_I,
\]
where three subscripts $0$, $R$, and $I$ indicate an unperturbed (free electron), Rashba spin splitting given in Eq.~(1) in the main text, and SL-SL interaction, respectively. Simply, $\mathcal{H}_0$ becomes $\hbar^2k^2/2m^*$ with $m^*$ an effective mass. For the matrix representation, we used four basis vectors $\ket{\mathrm{T},\uparrow}$, $\ket{\mathrm{T},\downarrow}$, $\ket{\mathrm{B},\uparrow}$, and $\ket{\mathrm{B},\downarrow}$ indicating states of electrons at top (T) and bottom (B) SLs with spin up ($\uparrow$) and down ($\downarrow$), respectively. 

Due to the oppositely-aligned local dipole moments at the top and bottom SLs, $\mathcal{H}_R$ can be represented by
\begin{align}
\bra{\mathrm{T},\uparrow}\mathcal{H}_R\ket{\mathrm{T},\uparrow}&=-\alpha_R k & 
\bra{\mathrm{T},\downarrow}\mathcal{H}_R\ket{\mathrm{T},\downarrow}&=\alpha_R k \nonumber \\
\bra{\mathrm{B},\uparrow}\mathcal{H}_R\ket{\mathrm{B},\uparrow}& =\alpha_R k & \bra{\mathrm{B},\downarrow}\mathcal{H}_R\ket{\mathrm{B},\downarrow}&=-\alpha_R k \nonumber 
\end{align}
where $\alpha_R$ and $k$ are Rashba strength parameter and crystal momentum.
On the other hand, $\mathcal{H}_I$ can be expanded, without losing generality, as
\[
  \mathcal{H}_I=E_I^0+E_I^1k^2+E_I^2k^4+\cdots.
\]
Since we consider the Hamiltonian for small $k$, the matrix elements of $\mathcal{H}_I$ becomes up to the 2nd order,
\[
 \bra{\mathrm{T},\uparrow}\mathcal{H}_I\ket{\mathrm{B},\uparrow}=E_I^0+E_I^1k^2, \qquad
 \bra{\mathrm{T},\downarrow}\mathcal{H}_I\ket{\mathrm{B},\downarrow}=E_I^0+E_I^1k^2.
\]
Note that the opposite spin states on the different SLs may not give repulsive SL-SL interactions, and thus $\braket{\mathrm{T},\uparrow|\mathcal{H}_I}{\mathrm{B}\downarrow}=0$, and so on. Thus, the total Hamiltonian ($\mathcal{H}$) becomes,
\begin{equation*}
\mathcal{H}=\mathcal{H}_0+\mathcal{H}_R+\mathcal{H}_I=
\begin{pmatrix}
\frac{\hbar^2k^2}{2m^*}-\alpha_R k & 0 & E_I^0+E_I^1k^2 & 0  \\
0 &  \frac{\hbar^2k^2}{2m^*}+\alpha_R k & 0 & E_I^0+E_I^1k^2  \\
E_I^0+E_I^1k^2 & 0 &  \frac{\hbar^2k^2}{2m^*}+\alpha_R k & 0  \\
0 & E_I^0+E_I^1k^2 & 0 & \frac{\hbar^2k^2}{2m^*}-\alpha_R k.
\end{pmatrix}
\end{equation*}
To obtain its energy eigenvalues, we solved its characteristic equation $|\mathcal{H}-E\mathcal{I}|=0$ to get 
\begin{equation}
\label{eigen_SI}
E_{\pm}=\frac{\hbar^2k^2}{2m^*}\pm\sqrt{\left(\alpha_R k\right)^2+\left(E_I^0+E_I^1k^2\right)^2},
\end{equation}
which are two doubly-degenerate solutions satisfying the degenerate condition guaranteed by the inversion symmetry. 
These two equations correctly reproduce two asymptotic band behaviors, such as Rashba-like bands for $\mathcal{H}_I\approx0$ and two parabolic bands for $\mathcal{H}_R\approx0$. The parameters in Eq.~(\ref{eigen_SI}) were determined by fitting the lowest conduction band (CB1) and second lowest one (CB2) computed by our first-principles calculation shown in Fig.~3 (a) in the main text as follow. From bands without SOC, $m^*$ and $E_I^1$ were first determined to be $m^*=0.15m_e$ and $E_I^1=12.83$~eV{\AA}$^2$, respectively. With these values fixed, $\alpha_R$ and $E_I^0$ were then determined for nonzero SOC cases and summarized in Fig.~3 (b) in the main text.

\begin{figure}[p]
\includegraphics[width=1.0\columnwidth]{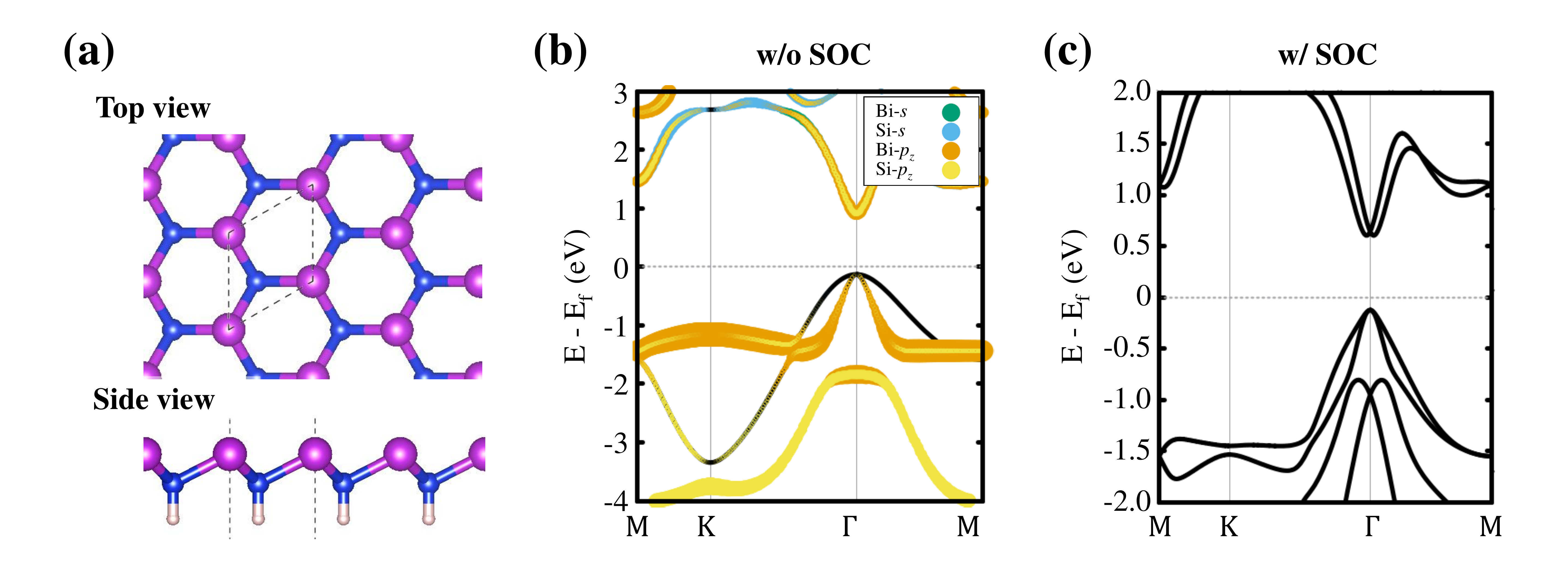}
\caption{(a) Top and side views of the crystal structure of 2D SiBi with same lattice constants with $\mathcal{I}$-Si$_2$Bi$_2$. The dangling bonds of Si are terminated by hydrogen atoms. The dashed lines indicate the primitive unit cell. The electronic structure of 2D SiBi is calculated by PBE functional without spin-orbit coupling (SOC) (b) and with SOC (c).
Bands in (b) were resolved into different orbitals as indicated in the inset. The line thickness indicates the degree of the orbital contribution.
\label{SiBi1}}
\end{figure}

\begin{figure}[p]
\includegraphics[width=1.0\columnwidth]{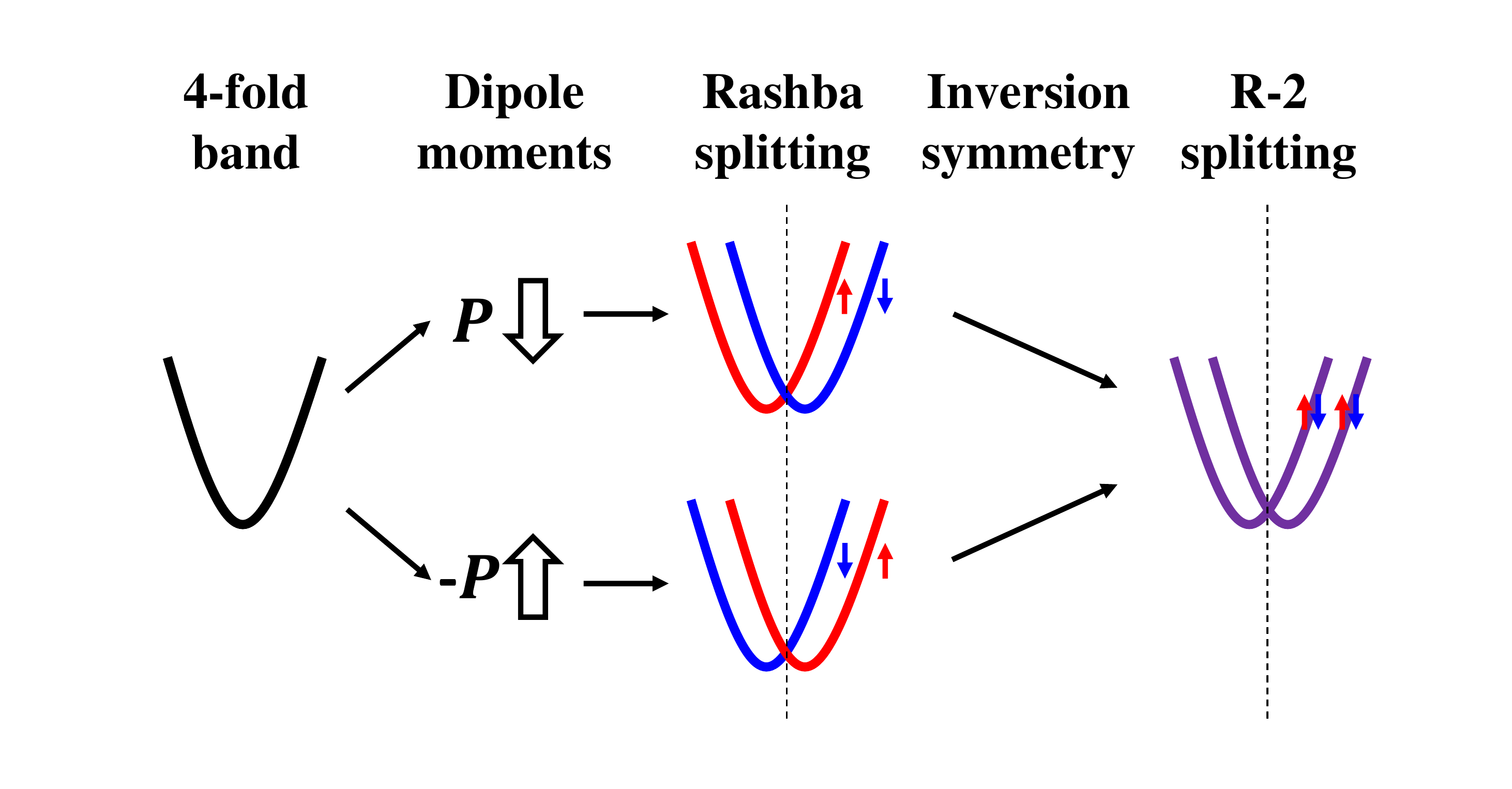}
\caption{Simple schematic band diagrams showing how R-2 spin splitting occurs without sublayer-sublayer (SL-SL) interaction. Without SOC, four bands from two spins and two SLs should be degenerate.
Since the local symmetry breaking generates a local dipole moment on one SL, which points oppositely to that on the other one, SOC induces the local Rashba spin splitting on each SL, depicted by red and blue band lines and arrows. The split spins are oppositely aligned due to the opposite dipole field directions. The inversion symmetry guarantees the degeneracy between the inversion partners, these bands should be combined into the purple band lines representing the hidden Rashba (R-2) spin splittings leading to the spatially-separated spin-up and down states or the spin layer locking (SLL).
\label{rashba_s}}
\end{figure}
\begin{figure}[p]
\includegraphics[width=1.0\columnwidth]{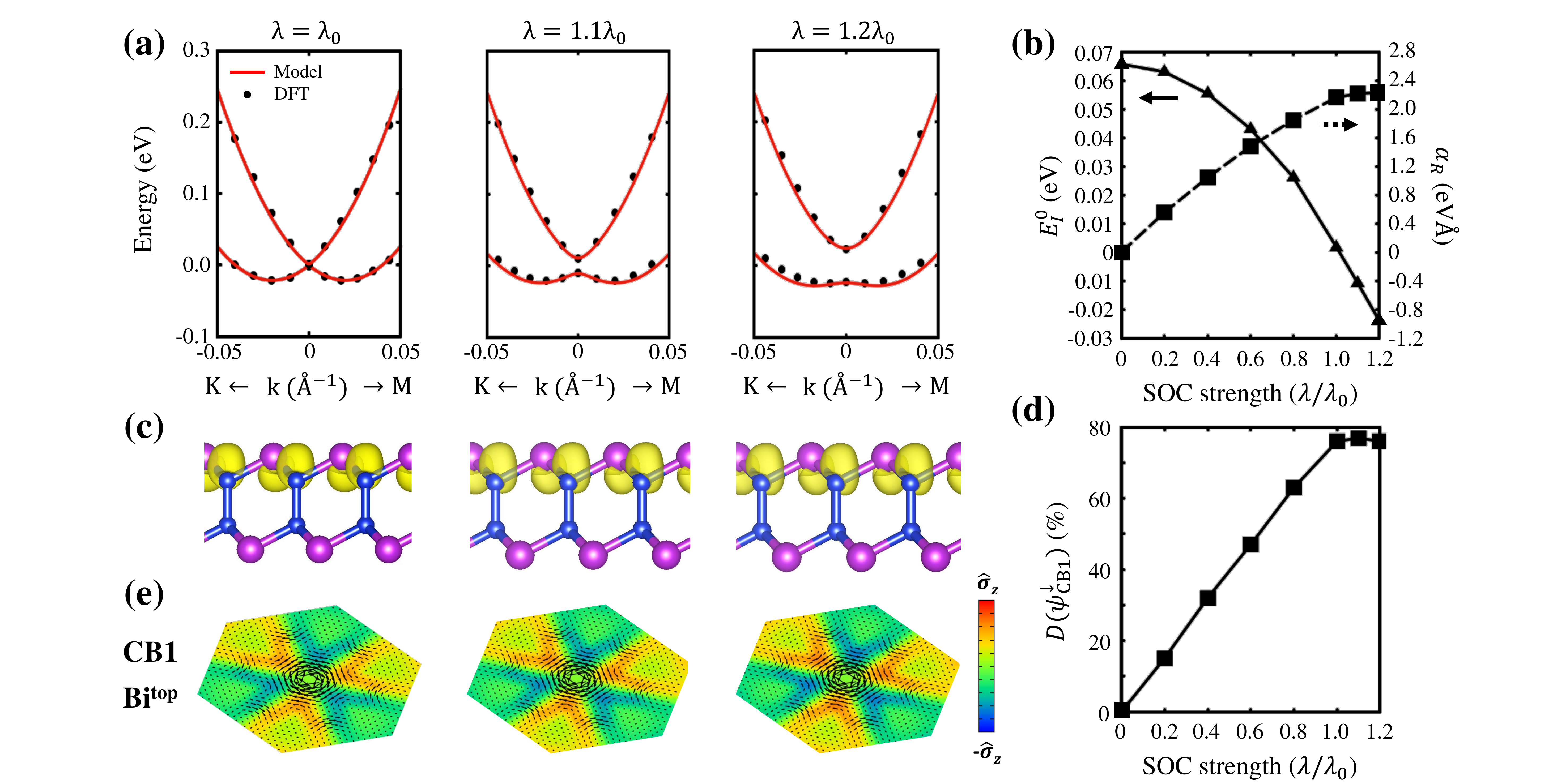}
\caption{Extension of Fig.~3 in the main manuscript: Evolution of the electronic structure with the SOC strength $\lambda/\lambda_0\in(1,1.2)$, where $\lambda_0$ is the real SOC strength: (a) two lowest conduction bands near the $\Gamma$ point. The black dots show the energy eigenvalues calculated by DFT, which were fitted by the model defined in the Eq.~(3) in the manuscript, plotted with red solid lines. (b) $\lambda$-dependence of the interaction energy ($E_I^0$) and the Rashba strength ($\alpha_R$) fitted in (a). (c) $|\psi^\downarrow_\mathrm{CB1}|^2$, spin-resolved wavefunction squared, calucated at $\mathbf{k_{\Gamma-M}}=(0.015,0) (2\pi/a)$ near the $\Gamma$ point. (d) Degree of wavefunction segregation D($\psi$) defined in Eq.~(4) in the manuscript evaluated from (c) as a function of $\lambda$. (e) Spatially-resolved spin maps on the top Bi atom layer plotted in the first Brillouin zone for the lowest conduction band. The size of black arrows and different colors indicate the in-plane and out-of-plane spin components
\label{further_soc_band}}
\end{figure}
\begin{figure}[p]
\includegraphics[width=1.0\columnwidth]{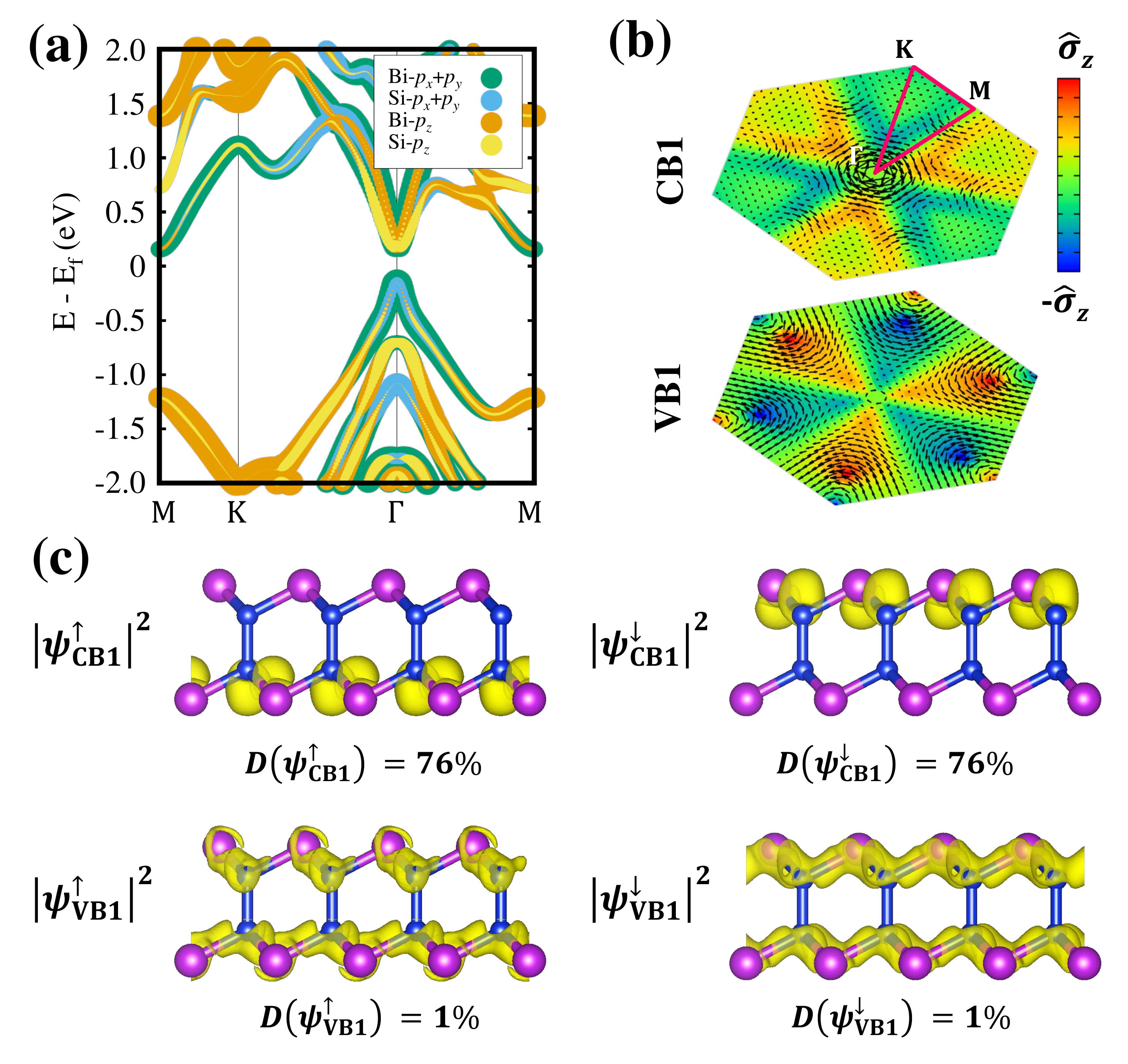}
\caption{
(a) Orbital resolved band structure of $\mathcal{I}$-Si$_2$Bi$_2$. As indicated in the inset, each color is assigned to each projected orbital and the line thickness indicates the degree of the orbital contribution.
(b) Spatially-resolved spin maps for the top Bi atom layer plotted in the first Brillouin zone for the lowest conduction band (CB1) and the highest valence band (VB1). The size of black arrows and different colors indicate the in-plane and out-of-plane spin components. 
(c) Real-space spinor wavefunction squared $|\psi^\sigma_{n}|^2$ calculated at $\mathbf{k}_{\Gamma-\mathrm{M}}=(0.015,0)(2\pi/a)$ near the $\Gamma$ point, with band index $n$ and spin index $\sigma$.
D($\psi$) defined in Eq.~(5) in the main text represents the degree of wavefunction segregation between the top and bottom SLs
\label{oam}}
\end{figure}

\begin{figure}[p]
\includegraphics[width=1.0\columnwidth]{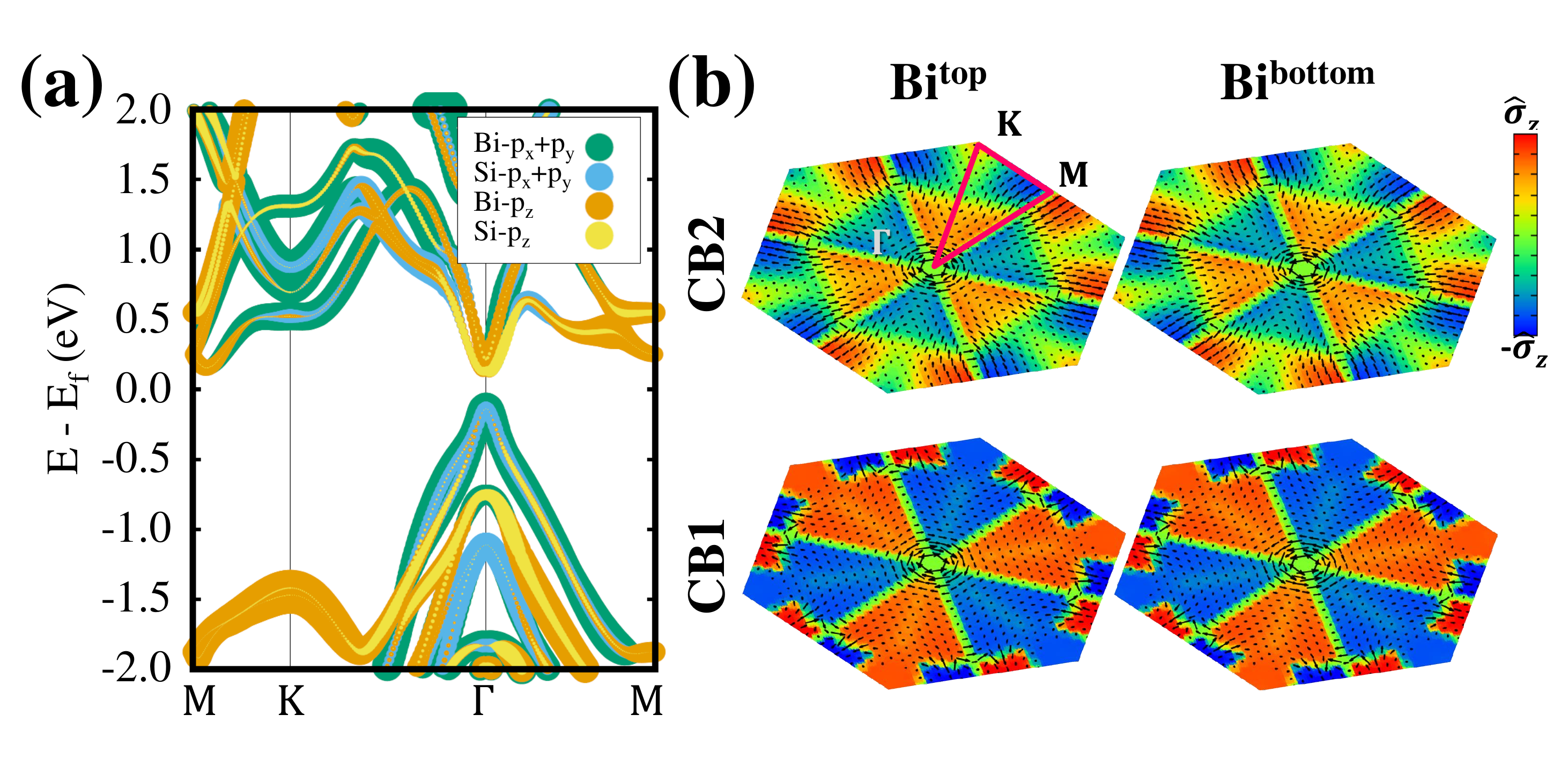}
\caption{The spin moment distribution for the CB1 and CB1 projected on Bi atoms in the top and bottom sublayers.
The hexagonal maps represent the first Brillouin zone of
$\mathcal{I}$-Si$_2$Bi$_2$, and the black arrows and color contour indicate
in-plane and out-of-plane spin-polarization components, respectively.
(a) Orbital resolved band structure of $\mathcal{M}$-Si$_2$Bi$_2$. As indicated in the inset, each color is assigned to each projected orbital and the line thickness indicates the degree of the orbital contribution.
(b) Spatially-resolved spin maps for the top and bottom Bi atom layers plotted in the first Brillouin zone for the lowest conduction band (CB1) and the second loweset one (CB2). The size of black arrows and different colors indicate the in-plane and out-of-plane spin components. 
\label{Rashba_M}}
\end{figure}

\bibliographystyle{apsrev}
\bibliography{my} 